\documentclass[twocolumn, 11pt]{aastex63}
\usepackage{epsfig,xcolor}
\usepackage{natbib}
\usepackage[normalem]{ulem}
\usepackage{comment}
\usepackage{mwe}

\usepackage{lineno}

\newcommand{\msolar}{\hbox{M$_\odot$}}
\newcommand{\muas}{\hbox{$\mu$\rm as}}
\newcommand{\omegacen}{\hbox{$\omega$~Cen}}
\newcommand{\highlightchange}[1]{{#1}}  
\begin{document}

\title{The Parallax of $\omega$ Centauri Measured from Gaia EDR3 and a Direct, Geometric Calibration of the Tip of the Red Giant Branch and the Hubble Constant}
\shorttitle{Parallax and TRGB Calibration of {\omegacen}}

\author{John Soltis}
\affiliation{Johns Hopkins University}

\author{Stefano Casertano}
\affiliation{Space Telescope Science Institute}

\author{Adam G. Riess}
\affiliation{Johns Hopkins University}
\affiliation{Space Telescope Science Institute}

\begin{abstract}
We use data from the ESA Gaia mission Early Data Release 3 (EDR3) to measure the trigonometric parallax of {\omegacen}, the first high precision parallax measurement for the most massive globular cluster in the Milky Way.  We use a combination of positional and high quality proper motion data from EDR3 to identify over 100,000 cluster members, of which 67,000 are in the magnitude and color range where EDR3 parallaxes are best calibrated.  We find the estimated parallax to be robust, demonstrating good control of systematics within the color-magnitude diagram of the cluster.  We find a parallax for the cluster of $0.191\pm0.001$ (statistical) $\pm0.004$ (systematic) mas (2.2\% total uncertainty) corresponding to a distance of $5.24\pm0.11$ kpc.  The parallax of {\omegacen} provides a unique opportunity to directly and geometrically calibrate the luminosity of the Tip of the Red Giant Branch (TRGB) because it is the only cluster with sufficient mass to provide enough red giant stars, more than 100 one magnitude below the tip, for a precise, model-free measurement of the tip.   Combined with the pre-existing and most widely-used measurements of the tip and foreground Milky Way extinction, we find  $M_{I,TRGB}=-3.97\pm0.06$ mag for the $I$-band luminosity of the blue edge.  Using the TRGB luminosity calibrated from the Gaia EDR3 parallax of {\omegacen} to calibrate the luminosity of SNIa results in a value for the Hubble constant of $H_0=72.1\pm2.0$ km s$^{-1}$ Mpc$^{-1}$.  We make the data for the stars in {\omegacen} available electronically and encourage independent analyses of the results presented here.

\end{abstract}

\section{Introduction}
\label{sec:intro}

The recent Early Data Release 3 \cite[EDR3;][]{Brown_2020, Lindegren_2020a, Riello_2020} from the Gaia mission \citep{Prusti_2016} opens a new chapter in the measurement of parallaxes, placing the precise and accurate determination of the distances to nearby Galactic globular clusters within reach. None may be more prized than that of {\omegacen}, the most massive cluster in the Galaxy at a distance of $\sim$ 5 kpc; a precise determination of its distance will characterize the luminosity of a broad sample of stellar types.  A precise distance to {\omegacen} will also provide a geometric calibration of the Tip of the Red Giant Branch (TRGB) feature, which in a cluster of this mass can be measured {\it directly} \citep{Bellazzini_2001} without the need for extrapolations or comparisons with other stellar features \citep{Capozzi_2020}.  The calibration of the TRGB is especially timely because it is a powerful primary distance indicator for evolved stellar populations that can be used to reach the hosts of Type Ia Supernovae (SNe Ia) and help determine the Hubble constant \citep{Freedman_2019, Yuan_2019, Jang_and_Lee_2015, Jang_Lee_2017, Jang_2021}; for a review, see \citet{Beaton_2018}. 

The Tip of the Red Giant Branch is the point of maximum brightness for Red Giant Branch (RGB) stars caused by the sudden start of helium fusion in low mass stars, a phenomenon known as the Helium Flash. After the flash, stars quickly expand and dim, which results in a visible decline in population in the color-magnitude diagram (CMD) at magnitudes brighter than the TRGB. The TRGB is an especially interesting feature of the CMD because of its usefulness as a distance estimator; the brightest red giants are bright enough to be seen in an interesting number of hosts to Type Ia SN, and the absolute magnitude of the TRGB in the $ I $ band, $ M_{I,TRGB} $, has only a modest dependence on metallicity and age of the underlying stellar population \citep{Lee_1993}.  Thus a calibration of $ M_{I,TRGB} $ yields the distance to other systems with an observed TRGB feature.  This has been done in the hosts of a number of type Ia supernovae to build a long-range distance ladder and to measure the Hubble constant \citep{Jang_Lee_2017, Freedman_2019}.  

Because the TRGB is not the identity of any individual star, but rather a feature (i.e., end point) of a distribution of stars, its luminosity has not been as easy to determine from trigonometric parallaxes as it is for individual stars that serve as standard candles such as Cepheid variables \citep{Benedict_2007, Riess_2018b} or RR Lyrae \citep{Benedict_2011}.  Rather, to calibrate the TRGB luminosity it has been necessary to determine the distance to an ensemble of stars at a common distance (although \citealt{mould_2019} show how a TRGB value could in principle be obtained from a sample of red giants at different distances using Gaia parallaxes); the resulting uncertainty includes that of the group's distance determination.  This approach has been utilized for the TRGB in the Large Magellanic Cloud (LMC) \citep{Freedman_2020}; however, while the LMC distance is accurately known \citep{Pietrzynski_2019}, the extinction internal to the LMC is substantial and subject to some disagreement in the literature  \citep{Freedman_2020, Nataf_2020, Yuan_2019, Jang_Lee_2017, Skowron_2020, Gorski_2020}.  As a consequence, the TRGB calibration in the LMC ranges  from $M_I=-3.95$ to $-4.05$ mag, corresponding to a 5\% range in the inferred value of $ H_0 $.  The halo of NGC 4258 provides another valuable target for TRGB calibration \citep{Macri_2006, Reid_2019, Jang_Lee_2017, Jang_2021, Mager_2008, Anand_2020}, yielding values between the middle and bright end of the range.  A calibration of the TRGB in globular clusters of the Milky Way with Gaia EDR3 offers some unique advantages to other TRGB hosts.  These include the direct use of trigonometric parallaxes (the ``Gold standard'' of geometric distance measurements), well-calibrated extinction estimates between us and the halo where Clusters reside \citep{Schlafly_2011}, negligible internal extinction in the system, and high photometric signal-to-noise for stars in the tip which, for a massive cluster, leaves little doubt about the location of the tip. 

However, calibrating the TRGB luminosity with Galactic globular clusters is subject to two challenges: mass and distance.  Bright red giants are relatively rare, and few clusters contain enough stars to fully populate the TRGB; \citet{Madore_1995} suggest that a robust measurement requires at least 100 stars present within 1 mag of the TRGB brightness.  The accuracy of parallax-based distances declines with increasing distance; an error of $ 5 \,\muas $  corresponds to 1\% in distance at 5~kpc, but 2\% at 10~kpc. {\omegacen} therefore provides the best opportunity in the Milky Way because it is relatively nearby, at a distance of about 5~kpc, and is the most massive Galactic cluster in \citet{Baumgardt_2018}, with an estimated mass of $ 3.55 \cdot 10^6 \, \msolar $. Indeed, \citet{Bellazzini_2001} find 185 RGB stars within one magnitude of their derived TRGB magnitude, nearly a factor of 2 better than the \citet{Madore_1995} recommendation.   Only three other clusters in the compilation by \citet{Baumgardt_2018} exceed $ 10^6 \, \msolar $, namely NGC~6388 ($ 1.06 \cdot 10^6 \, \msolar $), NGC~6441 ($ 1.23 \cdot 10^6 \, \msolar $), and NGC~6715 ($ 1.41 \cdot 10^6 \, \msolar $); all three, however, are estimated to be further than 10~kpc, and therefore their parallax calibration will be significantly more difficult.  Other potentially interesting clusters, such as M3, M13, and NGC 2808, have masses in the range $ 3\hbox{--}8 \cdot 10^5 \msolar $ and estimated distances between 6 and 10~kpc.

Several distance estimates have been published for {\omegacen} over the years using a wide range of methods which are discussed in detail in section 5.  In this work we examine the stars in {\omegacen} with astrometric and photometric information in the newly available Gaia EDR3 catalog, and find a precise direct parallax measurement for {\omegacen} yielding an accurate geometric calibration of its TRGB, taking into account known and possible systematic parallax errors and use this to calibrate the luminosity of the TRGB and the Hubble constant.  

The paper is organized as follows. Section 2 covers our data selection procedure, including our initial cluster membership criteria and quality cuts. Section 3 covers our secondary cluster membership selections and our distance measurement. In section 4 we report our TRGB calibration. Finally, in section 5 we discuss our results and their significance.

\section{Data Selection}
\label{sec:data}

The globular cluster {\omegacen} is extremely rich, reaches a very high density of stars in the center, and extends to a nominal tidal radius of about $ 57' $ \citep{Poveda_1975, Harris_1996}, although tidal tails have been found out to well over 1 degree \citep{Marconi_2014, Calamida_2017}, and \citet{FernandezTrincado_2015} have traced stars presumably stripped from {\omegacen} over a large fraction of the sky.  However, already at radii $ \sim 30' $, foreground/background Galactic stars have density comparable to cluster members \citep{Calamida_2017}, and obtaining a clean sample of cluster stars requires additional selections.  

We used a radius of $ 45' $ from the nominal cluster center (${\rm RA} = 210.697 $, $ \delta = -47.479 $; \citet{Eadie_2016a, Eadie_2016b}) for our initial extraction.  We selected stars from the Gaia archive using the following search:

\footnotesize\texttt{
\begin{tabbing}
\noindent \hspace{10pt}\=\hspace{10pt}\=\hspace{10pt}\=\kill
SELECT TOP 2000000 * \\
FROM gaiaedr3.gaia\_source \\
WHERE CONTAINS(POINT('ICRS',gaiaedr3.gaia\_source.ra,\\
\>\>    gaiaedr3.gaia\_source.dec), \\
\>   CIRCLE('ICRS', \\
\>\>    COORD1(EPOCH\_PROP\_POS(201.697,-47.479472,0,-3.2400,\\
\>\>\>     -6.7300,234.2800,2000,2016.0)), \\
\>\>    COORD2(EPOCH\_PROP\_POS(201.697,-47.479472,0,-3.2400,\\
\>\>\>     -6.7300,234.2800,2000,2016.0)), \\
\>\>    0.75) \\
)=1\\
\end{tabbing}
}
\normalsize

We also applied a series of quality cuts on the data, using histograms of the distributions of the parameters to make our selections. We choose data with: 
\texttt{
\begin{tabbing}
astrometric\_excess\_noise\_sig \= $ \leq 8$,\\
astrometric\_excess\_noise      \> $ < 4$,\\
soltype                         \> $ > 3$,\\
phot\_bp\_rp\_excess\_factor    \> $ < 2.5$,\\
phot\_proc\_mode                \> $ = 0$, {\rm and} \\
astrometric\_gof\_al            \> $ < 4$.\\
\end{tabbing}
}
After these quality cuts, our starting sample includes 178,548 stars.

\begin{figure}
    \centering\raggedright
    \resizebox{3in}{!}{\includegraphics{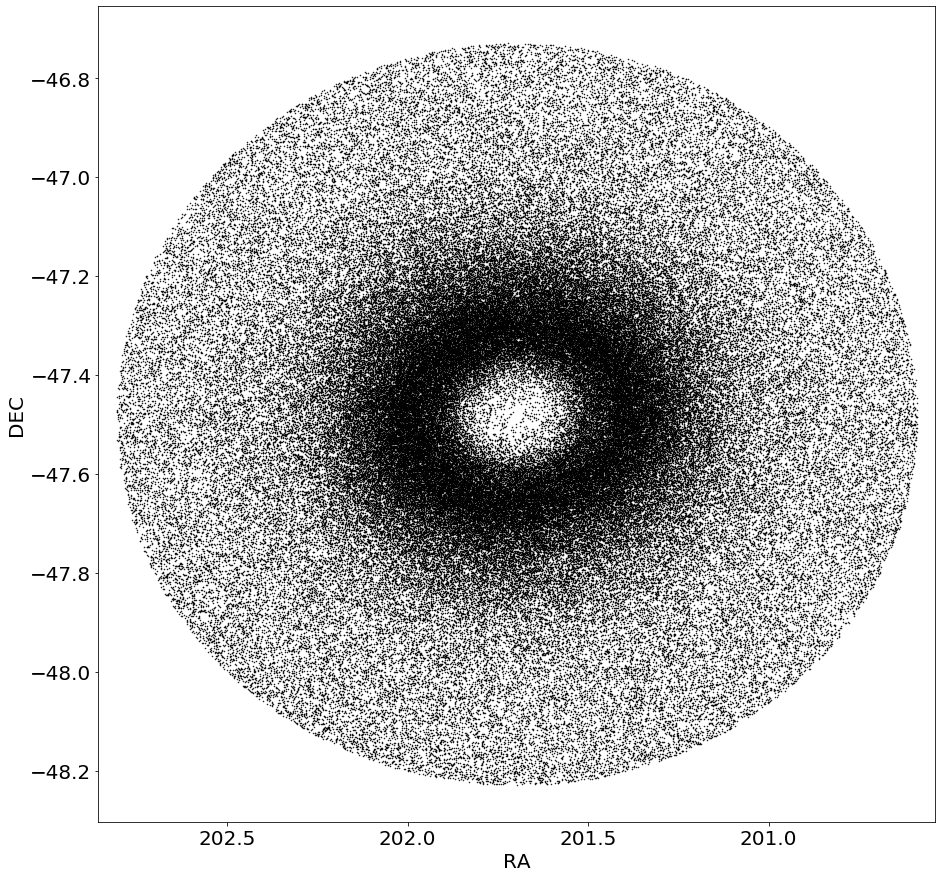}}
    \vspace{10pt}
    \resizebox{3in}{!}{\includegraphics{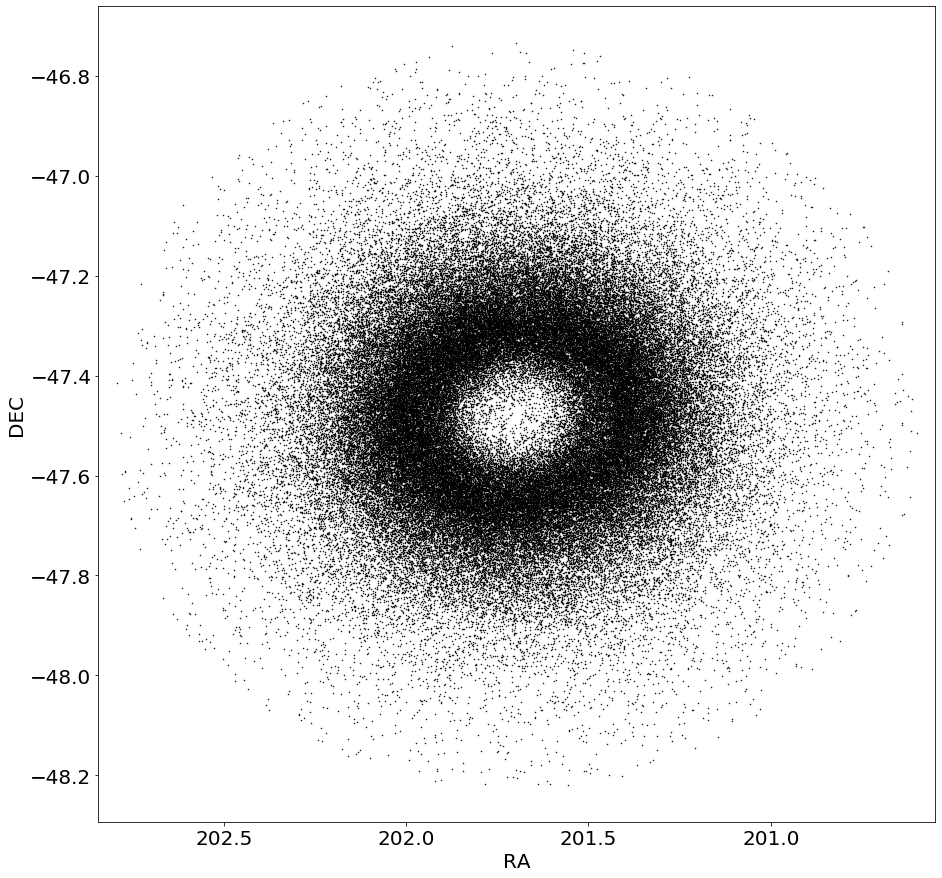}}
    \caption{\raggedright Distribution on the sky of sources selected from the EDR3 catalog.  Top: all sources within $ 45' $ of the nominal cluster center that pass our quality cuts.  The central region ($ 5' $ radius) is extremely crowded and is strongly depopulated in our initial sample as a consequence of our quality cuts.   Bottom: Presumed members according to the distance and proper motion selection shown in Fig.~\ref{fig:sep_ang_pm}.}
    \label{fig:good_sources}
\end{figure}

The central region of {\omegacen} is relatively crowded, challenging the quality of Gaia measurements in this region.
Figure~\ref{fig:good_sources} shows that after our EDR3 quality cuts, the central $ 5' $ radius of the cluster is essentially unpopulated.  From the perspective of a parallax estimate for the cluster, this is not a significant issue; cluster depth effects, discussed in Section~\ref{sec:cluster_depth}, are small, and have a modest dependence on angular distance from the center.  

\begin{figure}
    \centering
    \includegraphics[width=.45\textwidth]{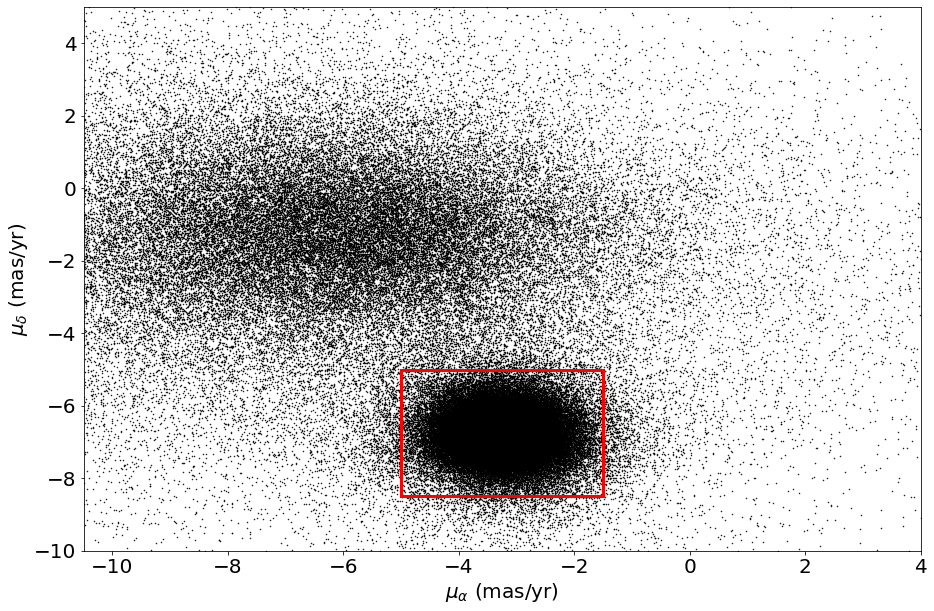}
    \caption{\raggedright The proper motions of the stars in our cone. A clear clustering corresponding to {\omegacen} is visible. The red box drawn is the region that was used to determine the nominal proper motion for the cluster. The other less dense cluster corresponds to Galactic field stars.}
    \label{fig:propmo_2d}
\end{figure}

The two-dimensional distribution of proper motions in right ascension and declination, $ \mu_\alpha $ and $\mu_\delta $, is shown in Figure~\ref{fig:propmo_2d}.  The proper motions for the selected sources fall into two separate components; the narrower component centered around ($-3.5, -6.5$) is due to cluster stars, while the much broader component roughly centered at ($ -6, -1 $) is due to field stars.  (Proper motions for field stars extend well beyond the range of the Figure.)  
The median proper motion for stars in the region ($-5 < \mu_{\alpha} < -1.5$, $-8.5 < \mu_{\delta} < -5$ mas/yr), shown by the red box in the Figure, is ($\mu_{\alpha},\mu_{\delta}$) = (-3.25, -6.76) mas/yr.  We \highlightchange{adopt} this as the nominal EDR3 proper motion for {\omegacen}, very close to the value $ (-3.24, -6.73) $ mas/yr reported by \citet{Baumgardt_2019} for Gaia DR2. \highlightchange{The uncertainty in the cluster proper motion is difficult to estimate, given the possibility of magnitude- and color-dependent offsets in proper motion (which correlates with parallax), and other correlated error terms.  We find that the proper motion varies by 0.03 mas/yr (rms) across magnitude bins, and adopt this value as a conservative estimate of relative systematic uncertainty. This uncertainty does not include the possibility of global offsets in the EDR3 proper motion for the cluster as a whole}. The proper motion width of the globular cluster component is likely due to a combination of internal motions and proper motion errors.  Internal motions are expected to contribute $ \sim 0.8 $ mas/yr per component at the center (assuming the parameters from \citealt{Baumgardt_2018}); \citet{Bellini_2018} measure an internal dispersion of $ 0.34 $~mas/yr just outside $ 15' $ from the cluster center.  Proper motion measurement errors have a median catalog value of 0.5--0.6 mas/yr (1D) for stars around $ G \sim 20 $.  

\subsection{Cluster Star Selection: Position and Proper Motion}

\begin{figure}
    \centering
    \includegraphics[width=.45\textwidth]{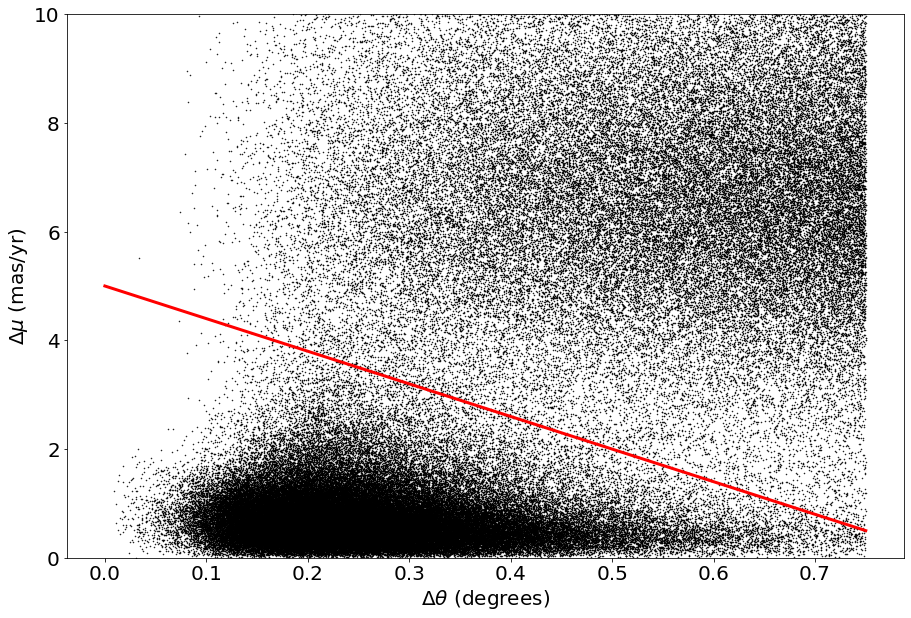}
    \caption{\raggedright The angular separation and proper motion difference of the stars from nominal are plotted. A clear clustering can be seen in the bottom left. All stars below the red line where assigned membership.}
    \label{fig:sep_ang_pm}
\end{figure}

We adopt a cluster membership criterion based on the combination of position and proper motion.  Figure~\ref{fig:sep_ang_pm} shows the distribution of the angular distance $ \Delta\theta $ of each star from the nominal cluster center, and the proper motion difference $\Delta \mu$ of each star from the cluster's motion determined above.  Stars obeying the relation:
\begin{equation}
    \Delta \mu < - 6 \frac{mas/yr}{degrees} \Delta \theta + 5 \frac{mas}{yr},
\end{equation}
\noindent i.e., below the red line in the Figure, are presumed to be cluster members.  Further refinements in membership on the basis of color and magnitude are discussed in the next Section.  After quality cuts and the joint angular separation /proper motion selection, the sample contains 108054 stars. 

\subsection {Cluster Star Selection: Magnitude and Color-magnitude Diagram}
\begin{figure}
  \centering
  \includegraphics[width=1\linewidth]{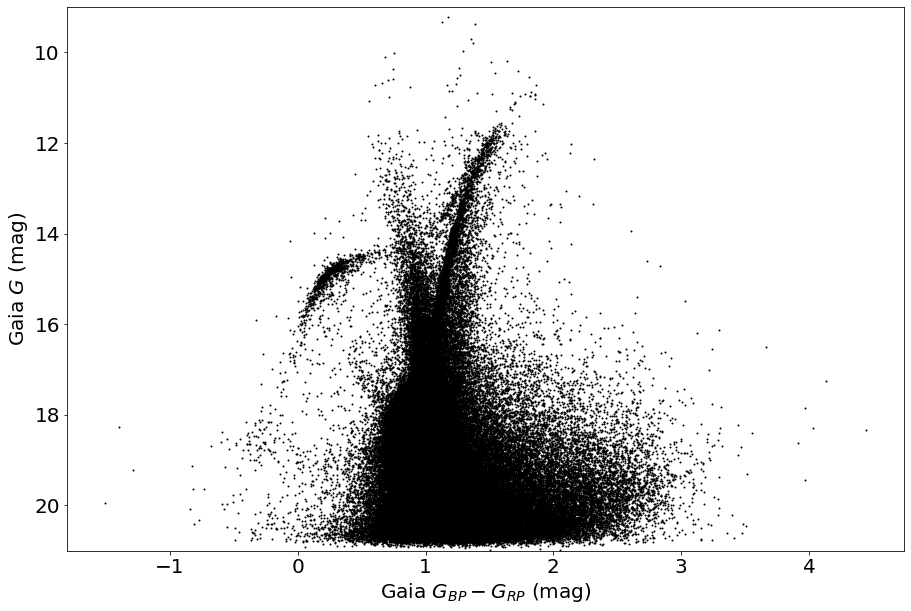}
  \noindent Pre-Membership Selection\par
  \vspace{20pt}
  \includegraphics[width=1\linewidth]{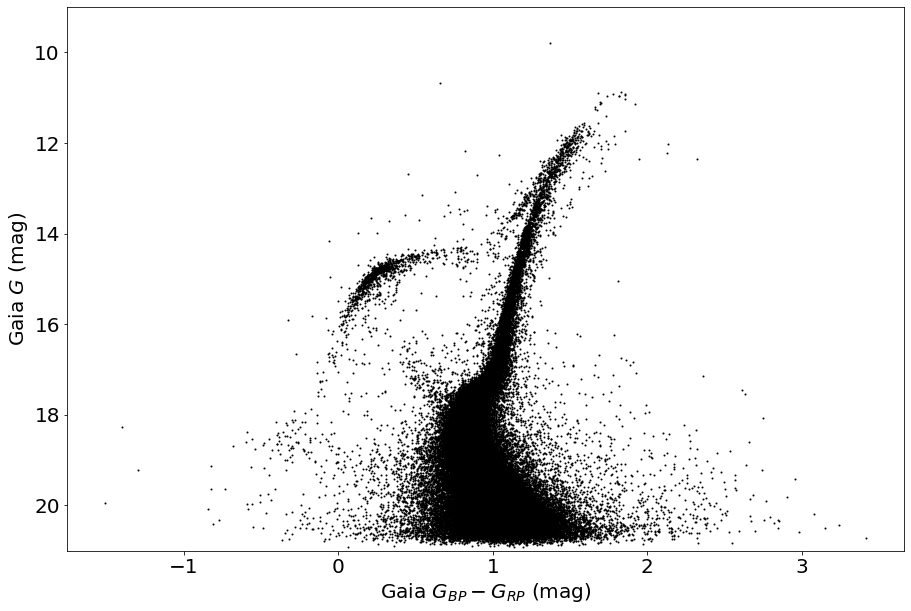}
  \noindent Post-Membership Selection\par
\caption{\raggedright Color-Magnitude diagrams of data pre-selection and post-selection. Removing contaminating stars sharpens the CMD.}
\label{fig:cmd_raw}
\end{figure}

Figure ~\ref{fig:cmd_raw} shows the distribution of stars in Gaia magnitude and color before (top) and after (bottom) the membership selection.  After the selection, the typical features of a globular cluster in color-magnitude space, namely main sequence, red giant branch, and horizontal branch emerge clearly, indicating that the vast majority of the remaining stars represent a well-defined, single-distance group. 

However, we can see in the plot a small number of apparent outliers in magnitude and color distribution.  These stars could be foreground or background stars, or stars with anomalous properties or EDR3 measurements.  Our goal is to determine a robust parallax for the system; given the large number of stars at our disposal, it is preferable to adopt a conservative approach and exclude potential contaminants, even at the cost of excluding a small number of true cluster stars.  

\begin{figure}
  \centering
  \includegraphics[width=1\linewidth]{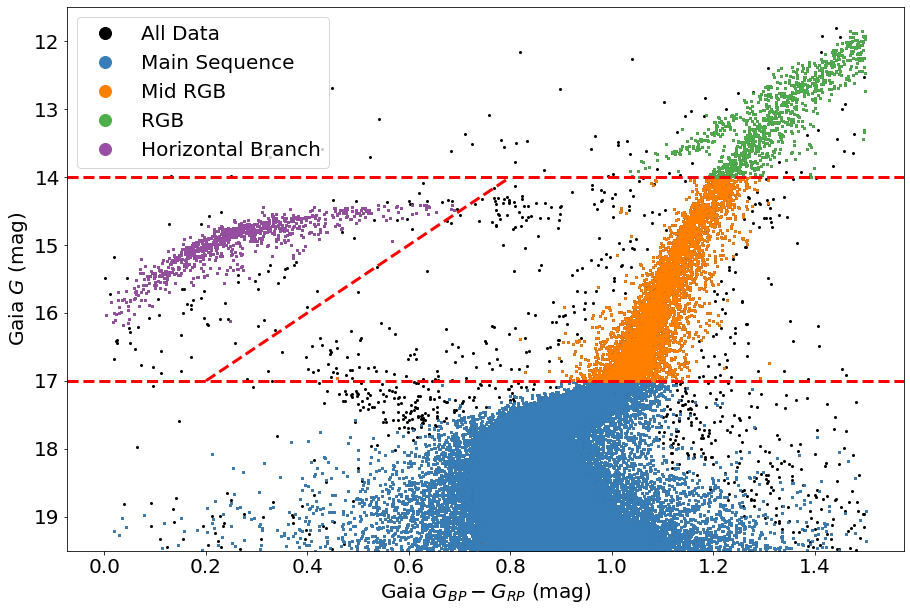}
\caption{\raggedright Color-magnitude diagram for {\omegacen}. The dashed red lines divide the CMD into the 4 regions used. Each region then had a cut imposed to remove outliers. See text for more details. The stars that survived the CMD cuts are plotted in color over top the pre-cut data.}
\label{fig:cmdregions}
\end{figure}

Therefore, we further restrict cluster membership using the color-magnitude diagram.  First we divide the color-magnitude diagram into 4 regions, as shown in Figure \ref{fig:cmdregions}.   Stars brighter than $ G = 11.5 $ or fainter than $ G = 19.5 $ are excluded, reducing the sample to 67347 objects.  We then bin these regions by $G$ magnitude; for each of these bins we determine a Gaussian kernel in color, and set a threshold kernel value.  Stars below the threshold value, shown in black in the Figure, are considered outliers and excluded from cluster membership.  This excludes an additional 880 stars, leaving a total of 66467 objects.  We could still expect a small fraction of contaminants, which are more easily addressed in the next section when we make use of parallax information.

\section{Distance Estimate}

 In EDR3, the characterization of the parallax offset has been greatly improved over that in DR2; \citet[hereafter L20]{Lindegren_2020b} provide an estimate of the parallax offset as a function of color, magnitude, and position on the sky, based on a combination of quasars, LMC stars, and physical pairs.  Indeed, the examination of the parallax of {\omegacen} stars across the CMD, spanning roughly 7 astronomical magnitudes and 1.5 magnitudes of color, provides a strong test of possible systematic uncertainties of the EDR3 parallaxes.  \highlightchange{Importantly, the EDR3 LMC star and quasar data provide the strongest constraints on the EDR3 parallax offset available and also fully cover the range of magnitudes, $12 < G < 19$ and $0 < B-R < 1.5$, for the stars we analyze in {\omegacen}.  (In contrast, the Cepheid stars analyzed in EDR3 by  \cite{Riess_2020} are in a brighter range, $6 < G < 10$, which have no overlap with LMC stars or quasars necessitating additional analysis of the magnitude-dependence of the offset in this range.)} All parallax values used in the remainder of this work are corrected according to the prescriptions of L20.

\begin{figure}
  \centering
  \includegraphics[width=1\linewidth]{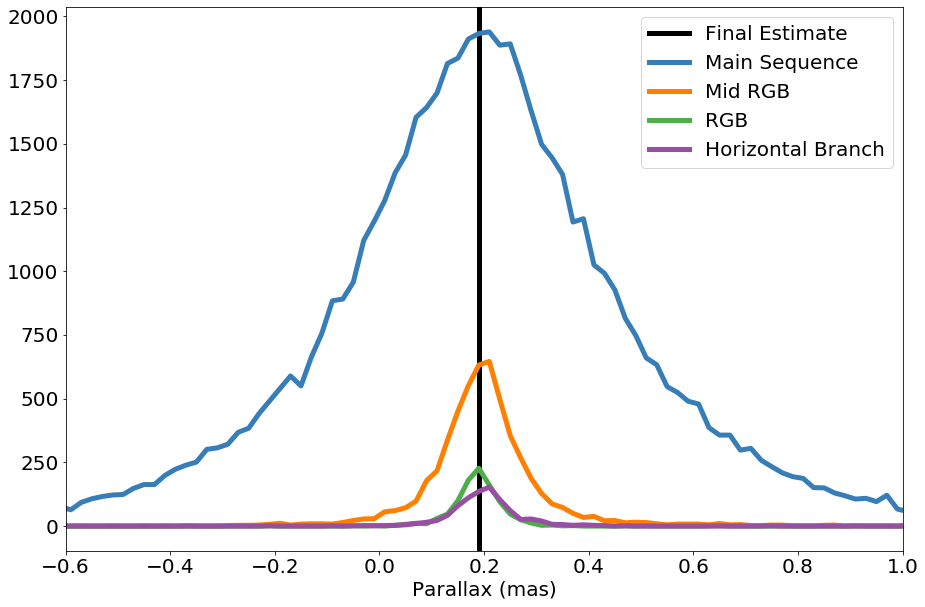}
\caption{\raggedright Histogram of parallax values selected by area in the color-magnitude diagram (as shown in Figure~\ref{fig:cmdregions}). Our final best estimate of the parallax for {\omegacen} is shown as a vertical black line.}
\label{fig:cmdhists}
\end{figure}

\begin{figure*}[b]
    \centering
    \null \vspace{-20pt}
    \includegraphics[width=6in]{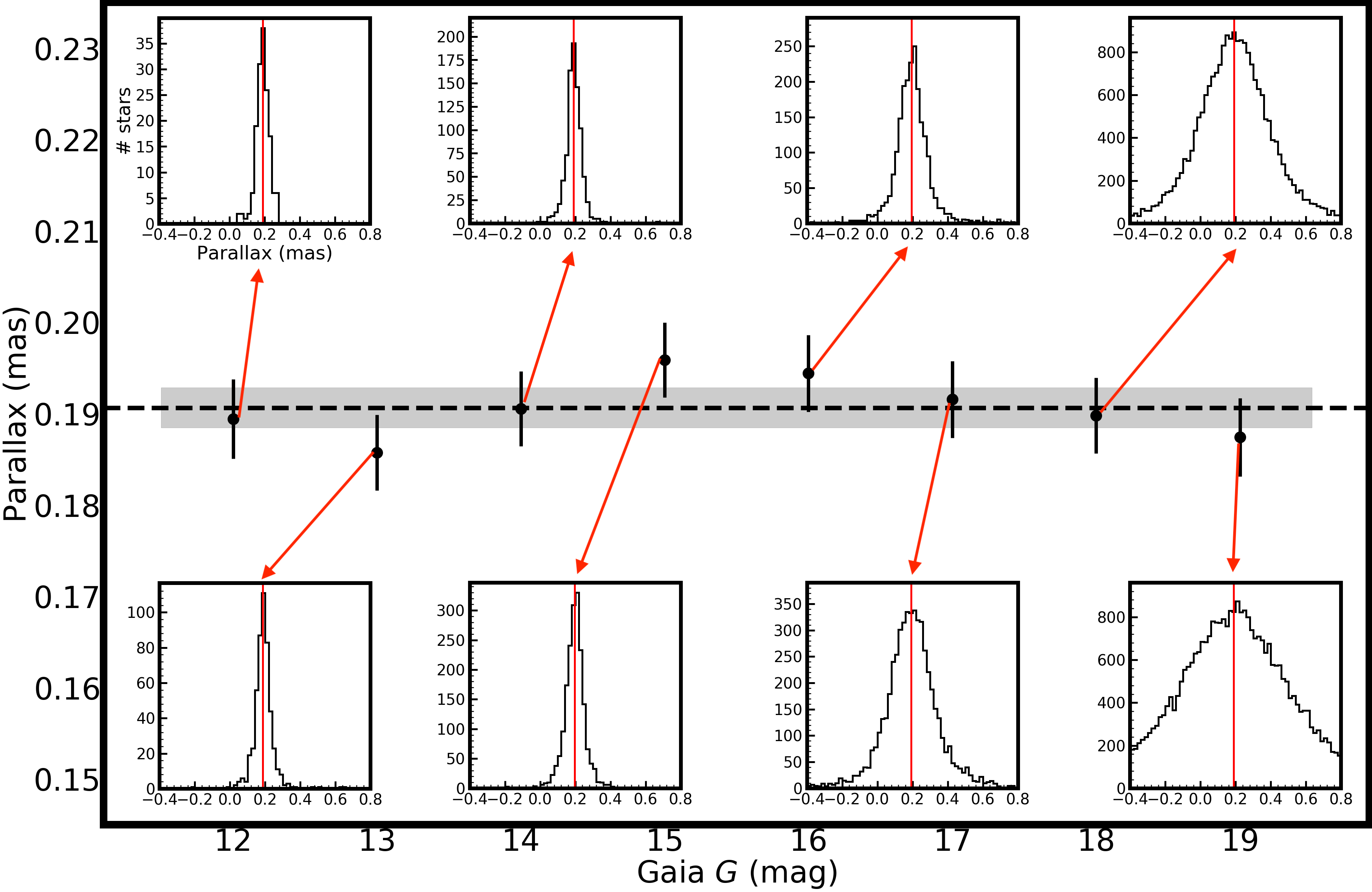}
    \caption{\raggedright Parallax binned as a function of magnitude. The dashed line is the mean parallax value of the magnitude bins as determined by bootstrapping (see text). The grey region contains 95\% of the bootstrap produce values. A histogram of each bin is shown, with the final weighted mean of the bin shown as a red vertical line. The error bars for each bin include a systematic term of 0.004~mas per bin.}
    \label{fig:magbins}
\end{figure*}

\begin{figure*}[b]
    \centering
    \null \vspace{-10pt}
    \includegraphics[width=6in]{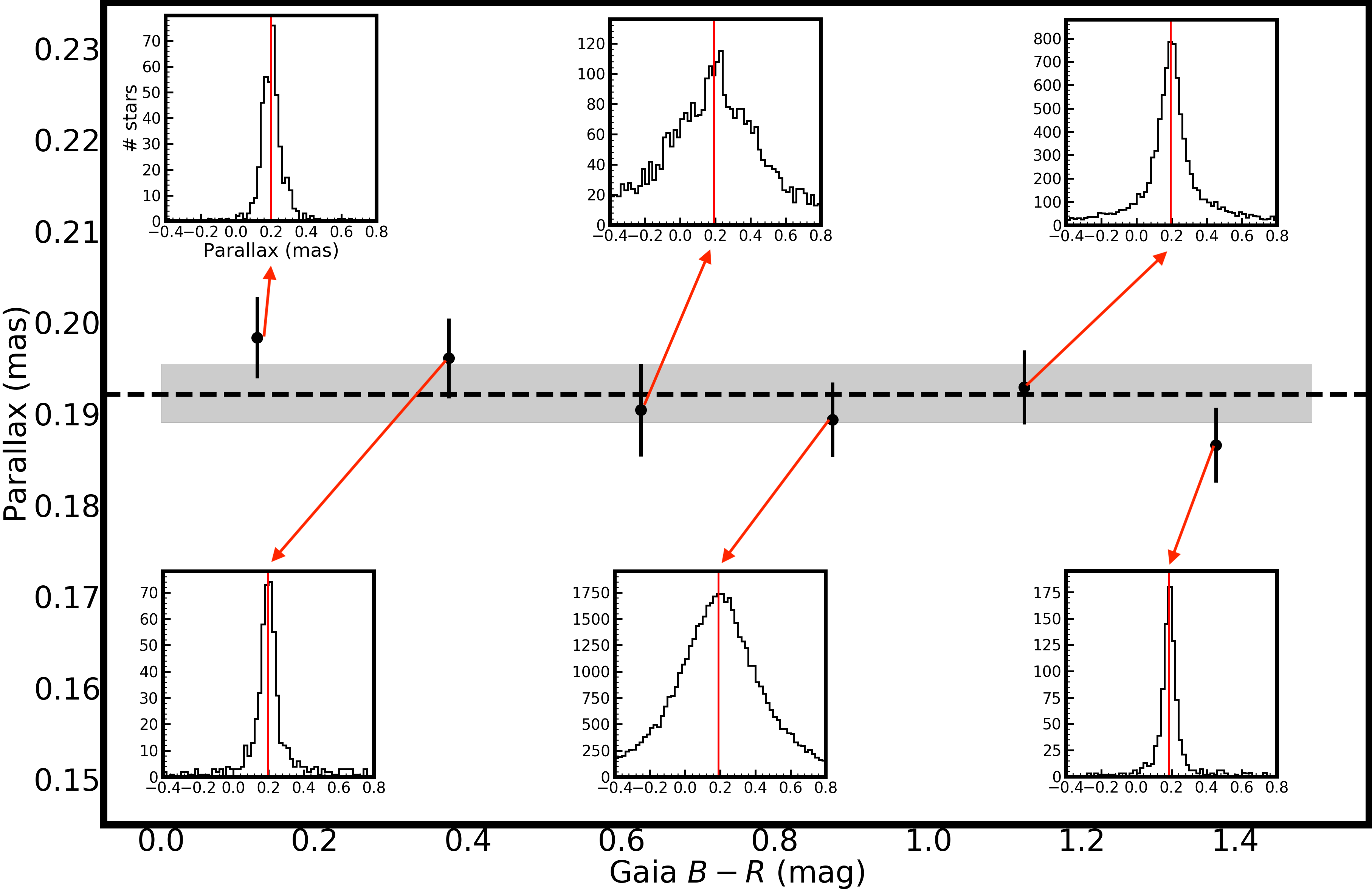}
    \caption{\raggedright Parallax binned as a function of color. The dashed line is the mean parallax value of the color bins as determined by bootstrapping  (see text).  The grey region is the 95\% confidence interval. A histogram of each bin is shown, with the weighted mean of the bin after sigma clipping shown as a red vertical line.  The error bars for each bin include a systematic term of 0.004~mas per bin.}
    \label{fig:colorbins}
\end{figure*}

A naive computation of the weighted mean parallax for the 66467 stars in the remaining member sample, based on catalog parallax errors, would lead to a nominal error of $ 0.4\,\muas $.  Because the systematic uncertainties in Gaia parallaxes are expected to depend on a star's magnitude and color, it is more appropriate to determine the cluster parallax for each region in the CMD, and also in fixed bins of $G$ magnitude and in fixed bins of $G_{BP} - G_{RP}$ color, so we can identify and characterize systematic uncertainties while simultaneously removing any remaining contaminants and outliers.  The histograms of parallax values in each of these groupings are shown in Figures~\ref{fig:cmdhists}, \ref{fig:magbins}, and~\ref{fig:colorbins}, grouped in terms of CMD regions, magnitude, and color, respectively; we choose a bin width of 1.0~mag in $ G $ and a color width of 0.25~mag in $G_{BP}-G_{RP} $.  At the bright end, each magnitude bin will contain a small number of stars with relatively high precision parallaxes, while at the faint end each bin will contain many more stars with individually less accurate parallaxes.  Within each bin, we determine the inverse-variance weighted mean parallax, and then apply an iterative 3 $\sigma$ rejection algorithm which includes the nominal parallax error so that rejection occurs for stars with individual values of $\chi^2 > 9$.  The parallaxes of each bin and their uncertainties are given in Table \ref{table:1}; for magnitude and color, they are also plotted, with their total uncertainties, in Figures~\ref{fig:magbins} and~\ref{fig:colorbins}.  As expected, the statistical uncertainty in each bin's parallax is fairly constant, as the change in the number of stars is balanced by the change in individual parallax precision.  These uncertainties range from 1 to 2~{\muas}.  

We provide the list of all the stars we analyzed (pre-CMD cut), 67347 in total, and their EDR3 parameters to simplify the replication of our results, \highlightchange{in the electronic form of this paper}, at \texttt{https://github.com/johnsoltis/Omega-Centauri-eDR3}, and upon request to the authors.

Perhaps our most important finding is that the parallaxes at this precision level do not trend with either color or magnitude, validating the quality of the relative characterization of the EDR3 parallax offset by L20 (at least at this location on the sky).  For comparison, we note L20 averaged the corrected EDR3 parallaxes over the full LMC and found the mean parallax thus derived agrees, within 1~{\muas}, with the detached eclipsing binary distance by \citet{Pietrzynski_2019}, which itself has an accuracy of 0.25~{\muas}, a meaningful validation of the ability to average EDR3 parallaxes over a small region.  Therefore we expect that, adopting the L20 correction, EDR3 parallaxes are good to $ \sim 1 $~{\muas} in absolute terms.

Nevertheless, we may conservatively expect some residual systematic uncertainty in the characterization of the parallax offset which depends on magnitude and color and L20 discusses this as having an expected uncertainty of "a few" {\muas} in the magnitude and color range where it is well-characterized.  We therefore adopt an uncertainty floor of 4~{\muas} in each of our magnitude and color bins, which we add in quadrature to the formal variance from the weighted average.  This assumption is likely on the conservative side, as the scatter from bin to bin shown in Figures 7 and 8 is somewhat smaller than the resulting uncertainty.  The error bars plotted in Figures~\ref{fig:magbins} and~\ref{fig:colorbins} include this systematic uncertainty.

Our statistical determination of the parallax of {\omegacen} is obtained by combining the nearly uniformly weighted values for each bin in either color or magnitude, and yields $ 0.1921 \pm 0.0017 $ and $ 0.1906 \pm 0.0011 $~mas, in good agreement with each other.  We adopt a value of $ 0.191 \pm 0.001 $~mas, corresponding to a distance modulus of $\mu=13.595 \pm 0.010 $~mag.

One final consideration is required concerning possible residual systematics.  Our sample of stars in {\omegacen} covers a region approximately $ 1^\circ $ across, a narrow region by Gaia standards, and parallax systematics are known to have some degree of angular covariance.  The analysis of the LMC in L20 is especially relevant for our purposes, since LMC stars, unlike quasars, occupy a similar range in magnitude and color as {\omegacen}; LMC stars are also sufficiently dense on the sky to permit a high quality analysis on smaller separations at constant parallax than is feasible with quasars.  Indeed, their Figure~14 shows a ``checkered'' pattern of residual zonal variations, with an amplitude of a few~{\muas}, and a correlation scale of $ \sim 1^\circ $.  Over the full LMC, several degrees across, these zonal residuals average down; however, our sample only covers roughly one correlation length, therefore systematics of a magnitude similar to the amplitude seen in the LMC should be expected.  From these LMC residuals in the L20 analysis we estimate the position-dependent systematic error for a patch the size of {\omegacen} to be 4~{\muas}.\footnote{Our estimate was based on EDR3 LMC data for sub-degree scales and the $\sim$ 2 {\muas} agreement between the EDR3 LMC parallax and its external value as an estimate to reach larger scales. \citet{Zinn_2021} estimates angular covariance of 10 \muas$^2$ between 0.3 degrees (size of \omegacen) and 10 degrees from asteroseismology of Red Giants and \citet{Maiz_2021} finds 40 \muas$^2$ between 10 degrees and larger scales from Quasars which combined provide an alternative uncertainty estimate of $\sim$ 7 \muas.  \citet{Maiz_2021} finds the same parallax for {\omegacen} that we find here.}  

Therefore our final estimate of the parallax of {\omegacen} is $ 0.191 \pm 0.001 $ (statistical) $ \pm 0.004 $ (systematic)~mas, or a final measurement of a distance modulus of $\mu=13.595 \pm 0.047$~mag. 

\begin{table}[h!]
\centering
\begin{tabular}{||c c c c c c||} 
 \hline
 Bin Range & \# Stars & $f_{\chi^2}$ & $\pi$ (mas) & $\sigma_{stat}$ & $\sigma_{tot}$ \\ [0.25ex] 
 \hline\hline
 11.5-12.5 & 156 & 0.929 & 0.1894 & 0.0017 & 0.0043\\ 
 12.5-13.5 & 496 & 0.889 & 0.1857 & 0.0011 &  0.0041\\
 13.5-14.5 & 882 & 0.947 & 0.1905 & 0.0009 &  0.0041\\
 14.5-15.5 & 1878 & 0.951 & 0.1959 & 0.0009 &  0.0041\\
 15.5-16.5 & 2142 & 0.943 & 0.1944 & 0.0013 &  0.0042\\
 16.5-17.5 & 5322 & 0.951 & 0.1915 & 0.0013 &  0.0042\\
 17.5-18.5 & 21690 & 0.967 & 0.1898 & 0.0011 &  0.0041\\
 18.5-19.5 & 33901 & 0.979 & 0.1874 & 0.0016 &  0.0043\\
\hline
$G$ Result & & & 0.1906 & & 0.0011\\
\hline
 0.0-0.25 & 432 & 0.961 & 0.1983 & 0.002 & 0.0045\\ 
 0.25-0.5 & 550 & 0.944 & 0.1961 & 0.0018 &  0.0044\\
 0.5-0.75 & 3370 & 0.973 & 0.1904 & 0.0031 &  0.0051\\
 0.75-1.0 & 51405 & 0.974 & 0.1893 & 0.0008 &  0.0041\\
 1.0-1.25 & 9767 & 0.951 & 0.1929 & 0.0006 &  0.004\\
 1.25-1.5 & 943 & 0.915 & 0.1865 & 0.009 &  0.0041\\[0.5ex] 
\hline
$G_{BP}-G_{RP}$ Result & & & 0.1921 & & 0.0017\\
\hline
 MS & 59263 & 0.973 & 0.1889 & 0.0008 & 0.0041\\ 
 Mid RGB & 5375 & 0.942 & 0.1933 & 0.0007 &  0.0041\\
 HB & 852 & 0.95 & 0.1967 & 0.0013 &  0.0042\\
 RGB & 977 & 0.909 & 0.1874 & 0.0008 &  0.0041\\[0.5ex] 
\hline
Regions Result & & & 0.1915 & & 0.0018\\
\hline
\end{tabular}
\raggedright
\caption{\raggedright Values for parallax binned by $G$ magnitude, $G_{BP} - G_{RP}$ color, and CMD region. $f_{\chi^2}$ is the fraction of stars that survive the 3 sigma clipping. $\sigma_{stat}$ is the statistical error and $\sigma_{tot}$ is the total error, which includes a 4~{\muas} systematic term.}
\label{table:1}
\end{table}

\subsection{Cluster Depth Effects}
\label{sec:cluster_depth}
The analysis carried out in this paper assumes that all stars in {\omegacen} have the same parallax.  In reality, the cluster has a small, but not insignificant extent along the line of sight; assuming rough spherical symmetry, an angular separation of $ 45' $ corresponds to a longitudinal extent of 1.3\%, or, for a parallax of 0.2~mas, about 0.003~mas.  This extent is significantly smaller than the formal parallax error of all stars we consider, and will act as a (negligible) additional dispersion in the effective parallax of selected members.  In the absence of extinction internal to the cluster, we expect that any incompleteness, primarily due to effects in projection on the sky, is essentially independent of position along the line-of-sight, and therefore averages down within each bin.  On the other hand, kinematic estimates are potentially affected by incompleteness, as velocity distribution, anisotropy, and projection effects depend both on the distance to the cluster center and, potentially, on the star mass (via its apparent luminosity).  Thus careful consideration of incompleteness effects will be required in the kinematic modeling of proper motion data from Gaia EDR3.

\section{Calibrating the TRGB}

The determination of the luminosity of the TRGB in the $I$-band is given in the simplest form as:
\begin{equation}
M_{I,TRGB}=m_I-A_I-\mu,
\end{equation}
\noindent where the distance modulus $\mu$ is obtained directly from the parallax, $m_I$ is the apparent brightness of the TRGB 
in the $I$ band, and $A_I$ is the line-of-sight extinction in the same band.

As discussed in Section~\ref{sec:intro}, {\omegacen} is ideal among Galactic globular clusters for the determination of $ m_I $.  It is the only globular cluster known to have at least 100 stars---185, according to \citet{Bellazzini_2001}---within one magnitude of the tip, a criterion for a robust measurement identified by \citet{Madore_1995}.  The cluster also has a low metallicity ([Fe/H]=-1.7 dex) and a relatively blue intrinsic color ($(V-I)_0$ $\sim 1.5$), making it an ideal calibrator of the TRGB stars in extragalactic halos \citep{Beaton_2019, Jang_Lee_2017}.  The color is within 0.1 mag of the fiducial TRGB color used in the characterization of \citet{Rizzi_2007}, and blueward of the $V-I < 1.9$ range suggested for the blue tip of Jang and Lee (2017).  

The most widely used study of the tip of {\omegacen} is from \citet{Bellazzini_2001}, hereafter B01, from the wide-field catalogue of \citet{Pancino_2000}.  B01 found:
\begin{equation}
    m_I = 9.84 \pm 0.04
\end{equation}

The determination was made with standard techniques \citep{Sakai_1996}, using an edge-detection or Sobel filter applied to a smoothed RGB luminosity function; the uncertainty was established through bootstrap resampling of the sample and is realistic for the number of stars and quality of the edge detection, and comports with simple error estimation of 0.03 mag based on a $\sim$ 0.2 mag width of a TRGB break and the existence of 10 stars within 0.1 mag of the break.  The value of $m_I$ is the same found by subsequent studies (e.g., \citealt{Bono_2008}). 

The extinction along the line of sight to {\omegacen} is due to the Milky Way disk, and occurs completely well in front of the cluster.  At a Galactic latitude of $ b \sim 15^\circ $, most of the extinction takes place well within 1~kpc along the line of sight.  Therefore the extinction to {\omegacen} may be taken as the full line of sight measured from the Galactic dust maps of \citet{Schlafly_2011} derived from the IRAS and COBE/DIRBE emission maps.  The size of the cluster is sufficient to average over sub-structure in Milky Way extinction so that the maps yield a robust result.  We estimate the uncertainty in the extinction using the prescription by \citet{Schlafly_2011}, i.e., 5\% of the extinction itself; see also the discussion in \citet{Brout_2019}.  The reddening estimate from \citet{Schlafly_2011} $E_{B-V}$=0.12 mag agrees within 0.5 $\sigma$ with the 0.13 mag of reddening measured by \citet{Bono_2019} for RR Lyrae stars in the cluster as well as the 0.11 mag measurement from \citet{Calamida_2005} using hot horizontal branch stars, falling between these two values.

Therefore:

\begin{equation}
    A_I = 0.215 \pm 0.011
\end{equation}

This results in $M_{I,TRGB}= -3.97 \pm 0.06$~mag.  According to \citet{Freedman_2019}, the color transformation to the filter $ F814W $ of the Hubble Space Telescope Advanced Cameras for Surveys subtracts 0.01~mag for TRGB stars, resulting in $M_{F814W,TRGB}=-3.98 \pm 0.06$ mag.  

This value is within the range of values of TRGB calibraions of $M_I$=-3.95 to -4.05~mag from the recent compilation of \citet{Capozzi_2020} and we discuss other calibrations further in the next section. The value measured here is fainter than the TRGB calibration used in recent determinations of the Hubble constant \citep{Freedman_2019} by 0.07 mag.  This change would raise the value of $H_0$ from \citet{Freedman_2019} by 3.2\%, yielding a determination of the Hubble constant from the TRGB-SN Ia distance ladder calibrated by the parallax of {\omegacen} from Gaia EDR3 of 72.1 $\pm 2.0$  km s$^{-1}$ Mpc $^{-1}$.  

\section{Discussion}

\subsection{Comparison with prior distance measurements for {\omegacen} }

The distance we derived from the Gaia EDR3 parallaxes of stars in the cluster is the same as that recently found from a kinematic measurement using Gaia DR2 by  \cite{Baumgardt_2019}. The recent determination of $ 5.24\pm 0.05 $~kpc by \citet{Baumgardt_2019}, in good agreement with the classical photometric distance of 5.2~kpc from \citet{Harris_1996}, is based on a comparison of the line-of-sight velocity dispersion profile against the DR2 proper motion measurements \citep{Brown_2018, Lindegren_2018}.  They obtain the line-of-sight velocity dispersion profile using the maximum likelihood approach briefly described in \citet{Baumgardt_2018}, and a proper motion dispersion profile from Gaia DR2 measurements with sufficiently small formal errors; they also use line-of-sight velocities and proper motions to exclude likely non-members.  They then match both line-of-sight and proper motion dispersion profiles to an array of N-body models, and thus determine the most likely distance to the cluster.  The error quoted by \citet{Baumgardt_2019} appears to be statistical only; a possible concern about systematics derives from the radially dependent selection effects, as the effective completeness limit for Gaia DR2 is a strong function of local density, and thus distance from the center.  The depletion of stars with good quality measurements from Gaia near the geometric center of the cluster may have some impact on kinematic studies of the cluster based on proper motions, since intrinsic velocity dispersion, anisotropy, and projection effects will depend on distance, true or projected, from the cluster center, and the mass distribution of EDR3 stars that pass our quality cuts can vary with selection effects. An earlier kinematic distance estimate by \citet{Watkins_2015}, using HST-based proper motions, puts {\omegacen} at a distance of $ 5.19^{+0.07}_{-0.08} $~kpc.  A distance estimate of $ 5.36 \pm 0.30 $~kpc was obtained by \citet{Thompson_2001} from the detached eclipsing binary OGLEGC~17 \citep{Kaluzny_1996}.   

A once challenging aspect of using Gaia parallaxes from the earlier DR2 release \citep{Brown_2018} to directly measure the parallax of {\omegacen} was the systematic uncertainty related to the parallax offset found by \citet{Lindegren_2018}, and which has been demonstrated to vary with source color, magnitude and position\cite[e.g.,][]{Riess_2018b, Arenou_2018, Zinn_2019}.  The DR2 offset was calibrated from quasars, which are bluer than the bulk of {\omegacen} stars and only cover a small fraction of their magnitude range.  The DR2 parallax estimates for {\omegacen} by \citet{Helmi_2018} and by \citet{Shao_2019} are $ 0.1237\pm 0.0011 $ and  $ 0.1368 \pm 0.0015 $~mas, respectively, and are substantially smaller than other determinations, presumably because the parallax offset estimated by \citet{Lindegren_2018} on the basis of quasar data is much smaller than the values found for brighter stars \citep{Riess_2018b, Arenou_2018, Zinn_2019}.   Our own analysis of the DR2 parallax data for stars in {\omegacen} revealed a large trend of parallax with star magnitude and color, leaving us to conclude that a parallax from DR2 was not reliable without further characterization of its parallax offset.  Thanks to the great improvement realized in EDR3, such a measurement is now feasible, and we exclude the DR2 parallax-based estimates from further consideration.

Both ground-based and HST observations have shown that {\omegacen} has a complex structure, with many separate color-magnitude sequences \citep{Bellini_2009b, Bellini_2010, Bellini_2017} that can have different structural \citep{Calamida_2017, Bellini_2017, Bellini_2018} and metallicity \citep{Milone_2017, Zennaro_2019} properties, including an unusually broad range in helium abundance for a globular cluster.  We would not expect this large range to influence the peak of the TRGB in the $ I $-band as the {\it mean} helium abundance in {\omegacen} is not much affected by the high and low sub populations, and the TRGB is known to be quite insensitive to star formation and chemical abundance in general \citep{McQuinn_2019}.  The complex kinematic structure of {\omegacen} does not affect its parallax measurement, but needs to be taken into account in kinematics-based distance measurements \citep{Watkins_2015, Baumgardt_2019}.  It is also worth recognizing that the distribution of abundances in the halos of galaxies where TRGB is often measured is largely unknown making comparisons difficult; it is not unlikely that the multiple populations of {\omegacen} could more closely resemble the complexity of a galaxy's stellar halo than a simpler, single-population system.
 
\subsection {Comparison with the TRGB calibration anchored to 47 Tuc}

The determination of the parallax to a globular cluster as massive as {\omegacen} represents a milestone of distance measurements and best means to {\it directly} calibrate the TRGB from a Globular cluster.  A more circuitous route was employed by \citet{Freedman_2020}, who used photometric distance offsets measured between clusters by \citet{DaCosta_1990} on the basis of the giant branch CMD, to bring 11 clusters to a common system with 47~Tuc, and formed a composite TRGB in the near infrared, anchored to 47~Tuc.  Each cluster is less massive than {\omegacen}, but together they provide enough stars to populate the TRGB.  They then used a mean detached eclipsing binary (DEB) distance to 47~Tuc of $\mu=13.27 \pm 0.07$~mag \citep{Thompson_2020} to obtain an absolute TRGB calibration, of $M_I=-4.05 \pm 0.10$ mag.  While consistent with the result we find here, the precision enabled from the Gaia EDR3 parallaxes suggests a revision faintward.  The \citet{Thompson_2020} distance for 47~Tuc is 0.03 mag farther than the parallax measured by \citet{Chen_2018} from Gaia DR2 by de-trending the previously discussed systematics in the parallax offset using nearby Small Magellanic Cloud (SMC) stars, and 0.05 mag farther than the value we find from EDR3 applying the same methods used here; with this distance, the composite Globular cluster analysis tied to 47~Tuc by \citet{Freedman_2020} would be revised to $M_I=-4.00 \pm 0.09$.

\subsection {Comparison with the LMC}

The distance to the LMC is known to high accuracy, thanks to the 20 late-type detached eclipsing binaries studied by \citet{Pietrzynski_2019}.  The LMC also has a well-populated TRGB, so that a very accurate empirical measurement of $ m_{I,TRGB} $ can be obtained \citep{Gorski_2018}.  However, the determination of the extinction {\it internal} to the LMC itself is more challenging. 

\citet{Skowron_2020} have recently published detailed and extensive maps of the LMC extinction based on OGLE data. 
The new maps calibrate the intrinsic color of Red Clump (RC) stars out to 5--8 degrees from the center, where the dust internal to the LMC is negligible, and the foreground reddening can be removed using the maps of \citet{Schlafly_2011}.  \citet{Skowron_2020} measure both the intrinsic RC color and its gradient (due to the metallicity gradient), a significant improvement over the maps of \citet{Haschke_2011} which assumed a fixed RC color of uncertain origin.  These maps can be used to correct the measurements of the TRGB across the LMC by \citet{Gorski_2018}, resulting in an accurate extinction-corrected measurement of $ m_I $.  The distance calibration of \citet{Pietrzynski_2019} then yields an absolute calibration of $ M_{I,TRGB} = -3.96 $~mag.  Zone-by-zone details are given in Table~II; the extinction varies by location, as expected, with a typical value $ A_I \sim 0.12 $~mag.  This calibration is in excellent agreement with the value we find here from {\omegacen}, but is $ \sim 0.1 $~mag fainter than the value from \cite{Freedman_2019}.  Some of the differences in the latter paper include a larger extinction $ A_I \sim 0.17 $, derived by F20 on the basis of TRGB colors, and a brighter value of $ m_I = 14.595 $ vs.~the value $ m_I=14.635 $ obtained by \citet{Gorski_2020}.  These two terms account for a difference by 0.09~mag.  Unfortunately we are unable to carry out a field-by-field comparison, as \citet{Freedman_2019} do not indicate which fields in the LMC were used in their analysis.  While the determination of the LMC extinction is challenging, it would be surprising if the extinction of stars near the TRGB was greater than those of RC star sight-lines, as TRGB stars should be older and further diluted; it is especially puzzling to note a difference as large as 0.04 mag for the apparent magnitude of the tip, since both groups use the same OGLE-III data, and the precision of the tip measurement has been claimed to be better than 0.01~mag in the analysis by \cite{Freedman_2019}.  This discrepancy warrants more detailed comparisons.

\subsection {Implications of the {\omegacen} measurement}

With new data from Gaia EDR3, the new extinction maps of the LMC (and SMC) from the OGLE Team, and new data from HST we may look forward to renewed efforts to calibrate and refine the luminosity of the TRGB.  The calibration of TRGB derived from the Gaia EDR3 parallax of {\omegacen} indicates a value of the Hubble constant of 72.1 $\pm 2.0$  km s$^{-1}$ Mpc $^{-1}$, in good agreement with other local measures, including those from Cepheids \citep{Freedman_2012, Riess_2020, Huang_2020}, although significantly larger than the value predicted by Planck CMB data used to calibrate $\Lambda$CDM (see \citet{Verde_2019} for a Review).  We may hope that additional work may shed new light on the oft-called ``Hubble Tension'' and lead to an explanation of its origin.

\begin{table*}[h!]
\centering
\begin{tabular}{||c c c c c c||} 
 \hline
 G18 Field & RA & DEC & $m_I$ & $A_I$ (Skowron) & $M_I$\\ [0.25ex] 
 \hline\hline
 LMC100 & 5:19:02.2 & -69:15:07 & 14.5810 & 0.0754071 & -3.97141\\ 
 LMC102 & 5:19:05.7 & -68:03:46 & 14.6610 & 0.118864 & -3.93486\\
 LMC103 & 5:19:02.9 & -69:50:26 & 14.6150 & 0.108667 & -3.97067\\
 LMC111 & 5:12:36.0 & -69:14:50 & 14.7140 & 0.141353 & -3.90435\\
 LMC112 & 5:12:21.5 & -69:50:21 & 14.6020 & 0.108817 & -3.98382\\
 LMC116 & 5:07:03.6 & -67:28:25 & 14.6820 & 0.102840 & -3.89784\\
 LMC120 & 5:05:39.8 & -69:50:28 & 14.6430 & 0.159487 & -3.99349\\
 LMC126 & 5:00:02.4 & -68:39:31 & 14.6200 & 0.111521 & -3.96852\\
 LMC127 & 4:59:33.6 & -69:14:54 & 14.6380 & 0.143930 & -3.98293\\
 LMC161 & 5:25:32.5 & -69:14:59 & 14.6240 & 0.112264 & -3.96526\\
 LMC162 & 5:25:43.3 & -69:50:24 & 14.5790 & 0.141297 & -4.03930\\
 LMC163 & 5:25:52.2 & -70:25:50 & 14.6480 & 0.110970 & -3.93997\\
 LMC169 & 5:32:22.8 & -69:50:26 & 14.6910 & 0.133787 & -3.91979\\
 LMC170 & 5:32:48.1 & -70:25:53 & 14.6000 & 0.102659 & -3.97966\\[0.5ex] 
\hline
Mean & & & 14.635 & 0.119 & -3.961 $\pm 0.011$\\
& & & & & (SD=0.04 mag)\\
\hline
\end{tabular}
\caption{\raggedright TRGB measurements from \citet{Gorski_2018} taken from across the LMC with updated extinction calculations from \citet{Skowron_2020} are shown. Absolute TRGB calibrations are then calculated using the DEB distance from \citet{Pietrzynski_2019}. The mean TRGB value is in good agreement with our {\omegacen} value.}
\null\vspace{-10pt}
\label{table:2}
\end{table*}

\section{Acknowledgments}

We thank David Nataf and Dorota Skowron for contributions toward understanding expectations of the LMC extinction. We are grateful to the entire Gaia collaboration for providing data and assistance which made this project possible. We congratulate them on their tremendous achievement to date.

\clearpage
\bibliographystyle{aasjournal}
\bibliography{myrefs}{}

\end{document}